\documentclass{IEEEtran}
\usepackage{graphicx}
\usepackage{dcolumn}
\usepackage{bm}
\usepackage{epsfig,wrapfig}
\usepackage{epstopdf} 
\usepackage{color}
\usepackage{siunitx}
\usepackage{textcomp}
\usepackage{cite}
\usepackage{algorithmic}
\usepackage{amsmath,amssymb,amsfonts}
\def\BibTeX{{\rm B\kern-.05em{\sc i\kern-.025em b}\kern-.08em
    T\kern-.1667em\lower.7ex\hbox{E}\kern-.125emX}}
\begin{document}

\title{\selectfont{An Impedance Surface Technique for \\Wideband Matching and Miniaturization of \\Circular Patch Antennas}}

\author{Mikhail Siganov, Stanislav Glybovski, and  Dmitry Tatarnikov

\thanks{M. Siganov, D. Tatarnikov, S. Glybovski are with the School of Physics and Engineering, ITMO University, St. Petersburg, 197101, Russia (e-mail: s.glybovski@metalab.ifmo.ru). 

This work was supported by the Ministry of Science and Higher Education of the Russian Federation  (Project FSER-2025-0009).

This work has been submitted to the IEEE for possible publication. Copyright may be transferred without notice, after which this version may no longer be accessible.}}

\maketitle

\begin{abstract}
In this work, we propose a technique to enhance the performance of circular patch antennas using an embedded system of cylindrical impedance surfaces. The technique utilizes a derived analytical model of a circular patch antenna containing an arbitrary number of coaxial impedance surfaces along with nonlinear optimization algorithms. This allows for the calculation of the radius and impedance of each surface to achieve the desired matching frequency band for given antenna dimensions. We demonstrate two applications of adding an optimal set of  impedance surfaces into a compact circular patch antenna, i.e. the expansion of the matching frequency band keeping constant antenna dimensions, and the miniaturization (height reduction) maintaining the constant bandwidth. Two corresponding versions of a cavity-backed circular patch antenna each having three impedance surfaces are synthesized. Practical implementations for both versions are designed and considered in full-wave numerical verification of analytically predicted properties. A comparison with the conventional method using multi-element microstrip matching circuits shows a benefit in radiation efficiency.
  
\end{abstract}


\section{Introduction}
\label{sec:introduction}

\IEEEPARstart{P}{atch} antennas are open resonators formed by metal patches of certain shapes located above a ground plane (a flat metal shield). The advantages of patch antennas include simplicity, compactness, low manufacturing costs, attainability of omnidirectional or cardioid radiation patterns, as well as compatibility with the printed circuit board (PCB) technology.
Antenna engineers often face with the task of ensuring impedance matching in a fixed frequency band following strict limitations to the antenna dimensions. To address this issue, various miniaturization techniques have been proposed. For \textit{circular patch antennas} studied in this work, miniaturization means decreasing the diameter $2a$ of a disc patch and/or the height $h$ of its location above the ground plane. Methods for reducing the resonant diameter of a  circular patch antenna include: filling the internal space of a patch resonator with a dielectric \cite{balanis}, connecting the patch to the ground plane with a set of capacitive loads along the radiating slots (so-called \textit{fence antennas}) \cite{tatarnikov}, creating cuts in the patch or in the ground plane (see e.g. in \cite{compact_annular_ring_embedded_circular, microstrip_patch_antenna_miniaturisation}). As the patch radius $a$ becomes small compared to the wavelength, the patch antenna characteristics approach the Chu-Harrington limit \cite{chu}: the relative matching bandwidth shrinks with miniaturization \cite{q_factor_bounds_for_microstrip_patch}. In particular, matching bandwidth is proportional to $h$.

There are various methods for maximizing the matching bandwidth of patch antennas with size limitations, which include: adding a passively coupled microstrip resonator \cite{broadband_with_parasitic_patches}, using an L-shaped feed  \cite{patch_antenna_L_feed}, creating cutouts in the ground plane \cite{microstrip_and_printed_antennas} or patch \cite{annular_ring_coupled_circular_patch, characteristic_mode_analysis_of_a_class_of_empirical__U_slot}, or using a stepped impedance resonator \cite{a_novel_differential_fed_patch_antenna}.
The other method consists in adding standalone capacitive or inductive loads into the resonator, realized as vertical conductors connecting the patch with the ground plane \cite{a_design_of_low, a_design_of_compact_ultrawideband}.
However, the implementation of the above methods generally causes changes in the current distribution across the patch and/or ground plane through the excitation of higher-order modes distorting the radiation pattern.

On the other hand, there are bandwidth improvement methods keeping the radiation pattern shape unchanged. The most popular one implies connecting a nonradiative multi-element matching circuit composed of distributed or lumped elements to the feed 
 of a patch antenna (see e.g. in \cite{microwave_filters_matching_networks}).
A decreased radiation efficiency due to losses in the circuit elements as well as significantly increased complexity and size of the printed implementation at a large number of elements are among the disadvantages of that method. There is a lack of systematic broadband impedance matching techniques that maximize the radiation efficiency of compact patch antennas.  

The field distribution inside a circular patch resonator can be effectively controlled using a set of identical loads arranged with a subwavelength period. Such a periodic structure of loads placed at a fixed distance from the antenna axis 
is equivalent to a continuous cylindrical \textit{impedance surface} with zero thickness \cite{Glybovski2014}. The distributed reactive impedance of the surface can significantly modify the frequency dependence of the antenna's input impedance without changing its radiation pattern thanks to preserving the initial angular field dependence of the selected mode (with azimuthal index $m=0$ for an omnidirectional pattern, or $m=1$ for a cardioid one) along the radiating slot ring. Advantageously, the presence of the structure can be analytically modeled using the \textit{averaged boundary conditions} method \cite{Glybovski2014}. Moreover, thanks to the validity of the single-mode approximation ($m=\text{const}$) the properties of the antenna can be predicted using an equivalent circuit approach \cite{characteristic_mode_analysis_of}. It is worth noting that analytically synthesised cylindrical periodic structures of loaded conductors have been used to form scattering field patterns \cite{sipus}, build high-gain antennas and cloaks \cite{eleftheriades}, achieve spatial field concentration \cite{extreme_localization_of_fields}, and, finally, improve the matching bandwidth of a circular patch antenna by closely tuning the resonant frequencies of two modes of the same $m$ but different radial indices \cite{design_and_analysis_of_a_low_Profile, characteristic_mode_analysis_of}. However, none of the existing works provides a systematic synthesis method for ensuring the desired frequency properties of circular patch antennas with given dimensions.

In this work, we derive an analytical model of a circular patch antenna with an arbitrary number of internal cylindrical impedance surfaces. Using this model, we demonstrate the potential to expand the matching frequency band for the given compact antenna dimensions, as well as to reduce the antenna height for the given diameter and matching bandwidth. The proposed method is numerically compared with one using multi-element microstrip matching circuits.

\section{Analytical Model}
\begin{figure}[t]
	\centering  
	\includegraphics[width=0.95\linewidth]{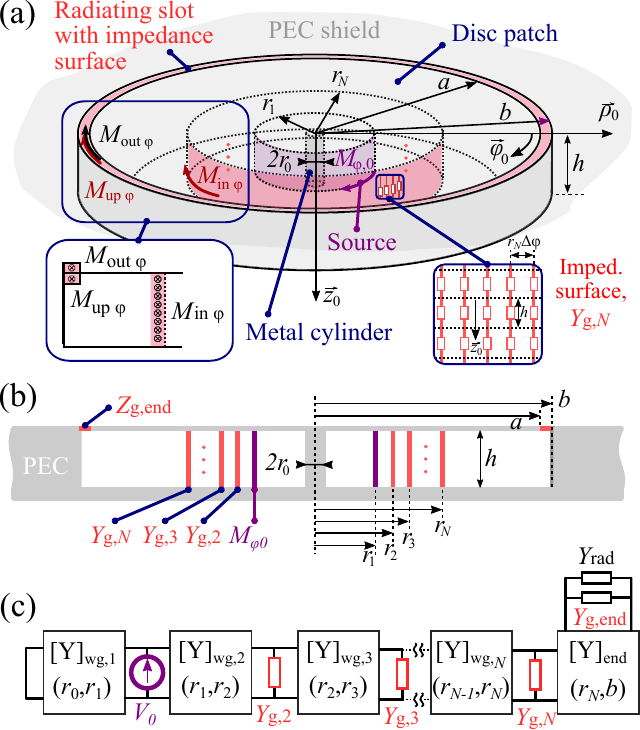}
	\caption{To the analytical model of a cavity-backed circular patch antenna with impedance surfaces, an annular radiating slot at $a<r<b$, and an infinite ground plane at $z=0$: (a) perspective view of the patch antenna model, showing one impedance surface at $r=r_N$; the inset on the left explains the choice of magnetic currents for a multi-mode description of the end block with the slot in the upper metal plane; the inset on the right schematically depicts a flat periodic structure of loaded conductors, which represents the practical implementation of an impedance surface; (b) cross-section of the patch antenna in the $\varphi=0$ plane; (c) the equivalent circuit of the antenna, which is a cascade connection of two-ports with a source and a load being the radiation admittance of the  slot.}
\label{schematic_analytics} \end{figure}

Let us consider the model of a cavity-backed circular patch antenna shown in Fig.~\ref{schematic_analytics}(a) \cite{cavity_backed_circular_patch_antenna}. The patch is a metal disc with a radius $a$ in the plane $z=0$. A resonator with metal walls forms a cavity with height $h$ and radius $b$ in an infinite metal shield, where the upper plane of the cavity coincides with the plane of the patch. This configuration creates an annular radiating gap in the plane at $a<r<b$. In the proposed method, $N-1$ coaxial cylindrical impedance surfaces are contained within the resonator, while an additional ($N$th) flat impedance surface is defined within the aperture of the radiating gap. Fig.~\ref{schematic_analytics}(a) shows only one cylindrical surface with radius $r_N$, while Fig.~\ref{schematic_analytics}(b) (cross-section) shows an arbitrary number of coaxial surfaces. Thus, the volume of the resonator $r<b$, $0 < z < h$ is divided into $N+1$ coaxial cavities by $N-1$ coaxial impedance surfaces located at  $r=r_2,r_3,..., r_N$, and the cylinder with a given surface magnetic azimuthal current (source) at $r=r_1$. Such a source models a slot-fed  (aperture-coupled) patch antenna with a thin coupling annular slot at $r=r_1$. 
A metal cylinder with a small radius $r_0$ is installed in the center of a patch antenna, connecting the patch to the bottom metal plane ($z=-h$). Each impedance surface with number $n=2,...,N$ is implemented in practice as a periodic structure of vertical conductors loaded with identical capacitive/inductive loads connecting the patch to the bottom plane. It is assumed that the conductors are arranged with a small period compared to the wavelength. For the impedance surface shown in Fig.~\ref{schematic_analytics}(a) the period of the conductors is equal to $r_{N} \Delta \varphi_N \ll \lambda$, where $\Delta \varphi_N$ is the angular period.

Let us assume the field of a TM type with the azimuthal index $m \ge 1$ corresponding to the angular field dependence of $e^{-i m\varphi}$. To calculate the characteristics of the patch antenna, each $n$th impedance surface is modeled as an infinitely thin cylindrical boundary described by impedance boundary conditions \cite{kontorovich1962coefficient,Glybovski2014}, which can be written as:
\begin{equation}
\begin{gathered}
\vec{E}(r = r_n+0) = \vec{E}(r = r_n-0) = \\ = Z_{\text{g}, n} \vec{n}_0 \times [\vec{H}(r = r_n+0) - \vec{H}(r = r_n-0)]\text{,}
\label{boundary_condition}
\end{gathered}
\end{equation}
where $\vec{n}_0$ is the normal to the surface, directed from the internal to the external region, and $Z_{\text{g},n}=Y_{\text{g},n}^{-1}$ is an imaginary macroscopic parameter called \textit{grid impedance} which depends 
on the microstructure of the loaded conductors. In other words, the tangential component of the averaged (over the period of the structure) electric field $\vec{E}$, which is continuous when passing in the radial direction through the boundary, is proportional to the difference in the averaged tangential component of the magnetic field $\vec{H}$. 
It should be noted that each coaxial layer bounded by adjacent coaxial cylindrical surfaces can be treated as a homogeneous section of a radial waveguide. In this single-mode approximation the field is constant along the $z$ axis as far as $h\ll \lambda$, while on each surface $r=r_n$ there is only one non-zero electric field component $E_z$ and one magnetic field component $H_{\varphi}$.

The grid impedance in boundary conditions (\ref{boundary_condition}) is calculated numerically or analytically for a flat infinite periodic structure in free space, consisting of parallel conductors with a step $r_n\Delta\varphi$, periodically loaded along the length by identical lumped elements with a step $h$ (see the inset on the right-hand part of Fig~\ref {schematic_analytics}(a)). The corresponding calculation methods are available, e.g., in Chapter 4 of \cite{tretyakov}. Once the grid impedance is known, it can be employed for solving various direct boundary problems in a relatively simple form. Moreover, using (\ref{boundary_condition}) it is possible to solve inverse boundary problems. Here we simplify the antenna synthesis problem by splitting it in two steps. On the first (macroscopic) step we determine the required grid impedance value and radius for each impedance surface, while on the second (microscopic) one we find the corresponding microstructures of the periodic loads to proceed to the practical implementation.

To find the characteristics of the antenna with $N$ different impedance surfaces shown in Fig.~\ref{schematic_analytics}(b) the decomposition approach \cite{peterson} is used. Accordingly, the internal space of the antenna is divided into individually described coaxial layers bounded by adjacent impedance surfaces (blocks). 
In the single-mode approximation, the field in each block is a sum of two TM$_{m, 0}$ waves with azimuthal order $m$ propagating in the positive and negative radial directions. To describe each individual block the electromagnetic equivalence principle can be employed. The impedance surfaces bounding each block in the radial direction can be replaced with ideally conducting cylindrical surfaces. Then it is possible to select surface magnetic currents $\vec{M}_{\text{S}} =\vec{\phi}_0 M_{\text{S},\varphi}= - \vec{n} _0 \times \vec{E}_z$ flowing along the surfaces in the azimuthal direction in such a way that they create the same fields inside the isolated block under consideration as in the full antenna. In the last formula  
$\vec{E}_z$ is the tangential component of the electric field on the surface, and $\vec{n}_0$ is the internal normal to the surface. Thanks to the single-mode approximation, it is possible to describe the block with number $n=2,3,...N$ as an equivalent two-port, the voltage at the terminals of which is equal to $V_{n-1}=hM_{\text{S},\varphi}( r=r_{n-1})$ on the left and $V_{n}=M_{\text{S},\varphi}(r=r_{n})h$ on the right, which is characterized by a matrix of linear parameters (for example, admittance matrix $[\text{Y}]_{\text{wg},n}$).
The combination of blocks is described by an equivalent cascaded connection of two-ports, shown in Fig.~\ref{schematic_analytics}(c).
In that cascade, impedance surfaces with grid impedance $Z_{\text{g},n}$ are replaced by lumped loads having equivalent admittance $Y_{\text{g},n}=Z_{\text{g},n}^{-1}$ connected in parallel to the terminals of the two-ports in the planes of their connection. We also define the excitation source of the antenna in a form of an azimuthal magnetic surface current $M_{\varphi, 0}$ on the surface $r=r_1$. In the cascade two-port circuit it is equivalent to the lumped voltage source $V_0=h M_{\varphi ,0}$ connected between the first and second blocks.

Consider calculating the input impedance of the antenna shown in Fig.~\ref{schematic_analytics}(b), containing $N-1$ coaxial surfaces inside the resonator, as well as an impedance boundary within the radiating slot ($N$ impedance surfaces in total). With this aim we first calculate the admittance matrix for each coaxial block, as well as for each impedance surface bounding the blocks by solving for the fields in the single-mode approximation. The admittance matrix $[\text{Y}]_{\text{end}}$ of the end block containing a radiating slot in the upper metal wall describes the interaction between azimuthal magnetic currents flowing along the internal boundary $r=r_N$ and magnetic currents within the radiating gap ($z=0$, $a<r<b$). To calculate $[\text{Y}]_{\text{end}}$, higher-order modes of a radial waveguide are taken into account when solving the excitation problem. Next, the transmission matrix of the cascade connection of two-ports connected to the right of the source is determined. To do this, the above admittance matrices are recalculated into transmission matrices, which are then multiplied.
The load of the cascade circuit is found as a 
 a parallel connection of the analytically calculated complex radiation admittance $Y_{\text{rad}}$ of the annular slot and the grid admittance $Y_{\text{g,end}}=Z_{\text{g,end}}^{-1}$ of the impedance surface filling the slot. Note that the corresponding angular distribution of the magnetic current within the slot (of azimuthal index $m$) is assumed.
Finally, to find the input impedance, we take into account the presence of a metal cylinder of radius $r_0$ in the center of the antenna by connecting a shorted section of a radial waveguide of length $r_1-r_0$ on the left of the source. The analytical expressions for the aforementioned admittance matrices are detailed in Appendix~A.

The equivalent-circuit model can be used for predicting the frequency-dependent behavior of the calculated input impedance for a given set of parameters. Alternatively, the same model can be used to determine the antenna parameters providing the desired frequency properties of a circular patch. This implies the first (macroscopic) step of the synthesis problem, followed by the determination of the required microstructures of the impedance surfaces at the second (microscopic) step. The macroscopic step is illustrated for two different optimization goals in the next section.

\section{Antenna Optimization for Broadband Matching and Miniaturization} 

To illustrate the application of the proposed analytical  model, here we study the possibilities to optimize the frequency properties of circular patch antennas with given electrically small dimensions by employing certain number $N$ of impedance surfaces.

We arbitrary select dimensions $a=25$~mm, $b=27$~mm for an antenna operating in the GNSS L1 band (1535--1610~MHz). We will consider the case $m=1$ (radiation with a maximum of the radiation pattern in the upward direction). Note that $2b\approx0.29\lambda_0$, where $\lambda_0$ is a free space wavelength at the central frequency $f_0$ of the band. The initial value of the height before miniaturization is $h=8$~mm. As a requirement for the matching level, throughout this paper we choose $|S_{11}|<0.31\approx -10$~dB, assuming a 50-ohm line impedance. The parameters to be optimized are $r_n$, $Z_{\text{g},n}$, where $n=2,...,N$ , $r_0$, $r_1$ and $Y_{\text{g,end}}$ ($2N+1$ parameters in total). 
Optimization is performed by iteratively increasing the number of impedance surfaces $N$, starting from the case $N=1$. 

In the case $N=1$ (referred hereinafter as the \textit{standard
antenna}), a single impedance surface is installed within the radiating slot (without internal impedance surfaces), while its grid admittance $Y_{\text{g,end}}=i\omega C_{\text{g,end}}$ can be adjusted to tune the resonant frequency of the TM$_{1,1}$ mode to $f_0$ (without the possibility of changing the bandwidth). This tuning method is equivalent (with fixed antenna dimensions, in terms of the bandwidth) to filling the resonator space with a solid dielectric of a certain dielectric constant \cite{dvtat_gps_satellite_surveying}. Conventionally, the source position $r_1$ can be adjusted for matching, and $r_0$ should be minimal. A practical realization of such an end-load impedance surface is based on a fence of capacitive plates distributed along the periphery of the disc patch. Advantageously, the fence of periodic metal plates that are low-weight and easy to fabricate can be realized even for relatively large $h$ up to 10--20 mm without using a solid bulky dielectric layer. In this work, we extend this approach also to internal impedance surfaces with $N>1$.

To obtain a patch antenna with improved frequency characteristics, we choose the following integral as the  goal function to be minimized:
\begin{equation}
\label{goal_function}
\int_{f_0-\Delta/2}^{f_0+\Delta/2} (|S_{11}|-0.31)^2 df \text{,}    
\end{equation}
where $\Delta=\text{1610-1535 MHz}$ is the GNSS L1 band width. Since this function has many minima in the space of varied parameters, the search for an antenna configuration with the largest matching band at given $h$ and $N$ requires the use of global optimization methods. Thus, in this work we used a genetic algorithm implemented in the Global Optimization Toolbox package of the Matlab commercially available software \cite{mathworksGeneticAlgorithm}. Note that for impedance surfaces the values of $C_n$ (capacitance) or $L_n$ (inductance) are introduced and used as the varied parameters to correctly describe the frequency behavior of $Z_{\text{g},n}=1/i\omega C_{n}$ or $Z_{\text{g},n}=i\omega L_{n}$, correspondingly.

\begin{table}
\label{underground_patch}
\centering
\caption{Optimal parameters of circular patch antennas with different numbers of impedance surfaces}
\begin{tabular}{l c c c c | c}   
\hline
$N$ & 1 & 2 & 3 & 4 & 3  \\ 
\hline
$h$, mm & 8.0 & 8.0 & 8.0 & 8.0 & 4.8 \\
\hline
\multicolumn{6}{c}{Varied parameters:}\\
\hline
$r_0$, mm & 0.1 &0.1& 0.1 &0.4& 0.1  \\ 
$r_1$, mm & 7.9  & 7.7  & 2.0  & 1.5 & 2.0 \\
$C_2$, nF & - & 0.1 & 5.5  &8.3 & 5.5  \\ 
$r_2$, mm & - & 19 & 3.2  & 2.4 & 3.2 \\
$L_3$, nH & - & - & 1.9 & 1.6 & 1.8 \\
$r_3$, mm & - & - & 8.1 & 4.6 & 8.2 \\
$C_4$, nF & - & - & - & 1.1 & -  \\
$r_4$, mm & - & - & - & 6.6 & - \\
$C_\text{end}$, pF & 77 &  65 & 91 &  78 & 137  \\
\hline
 Bandwidth, MHz& 32  & 30  & 63  & 68 & 32  \\
 \hline
\end{tabular}
\label{comparison}
\end{table}

The optimization results and the obtained parameter values are summarized in Table~\ref{comparison} for $h=8$~mm and $N$ ranging from 1 to 4. As can be seen from the results, the presence of two impedance surfaces does not allow the bandwidth to be expanded compared to the standard antenna. Adding a third surface allows one to approximately double the frequency band. It was shown that a further increase in $N$ does not lead to a significant bandwidth improvement (in particular, for $N=4$ the band is increased by only 8\% compared to the case of $N=3$). At the same time, the increase in $N$ with an electrically small antenna size significantly complicates the practical design of the antenna. Therefore, the compromising case $N=3$ was chosen for the second (microscopic) synthesis step and numerical simulations of a practical antenna design.

\begin{figure}[t]
	\centering  
	\includegraphics[width=0.95\linewidth]{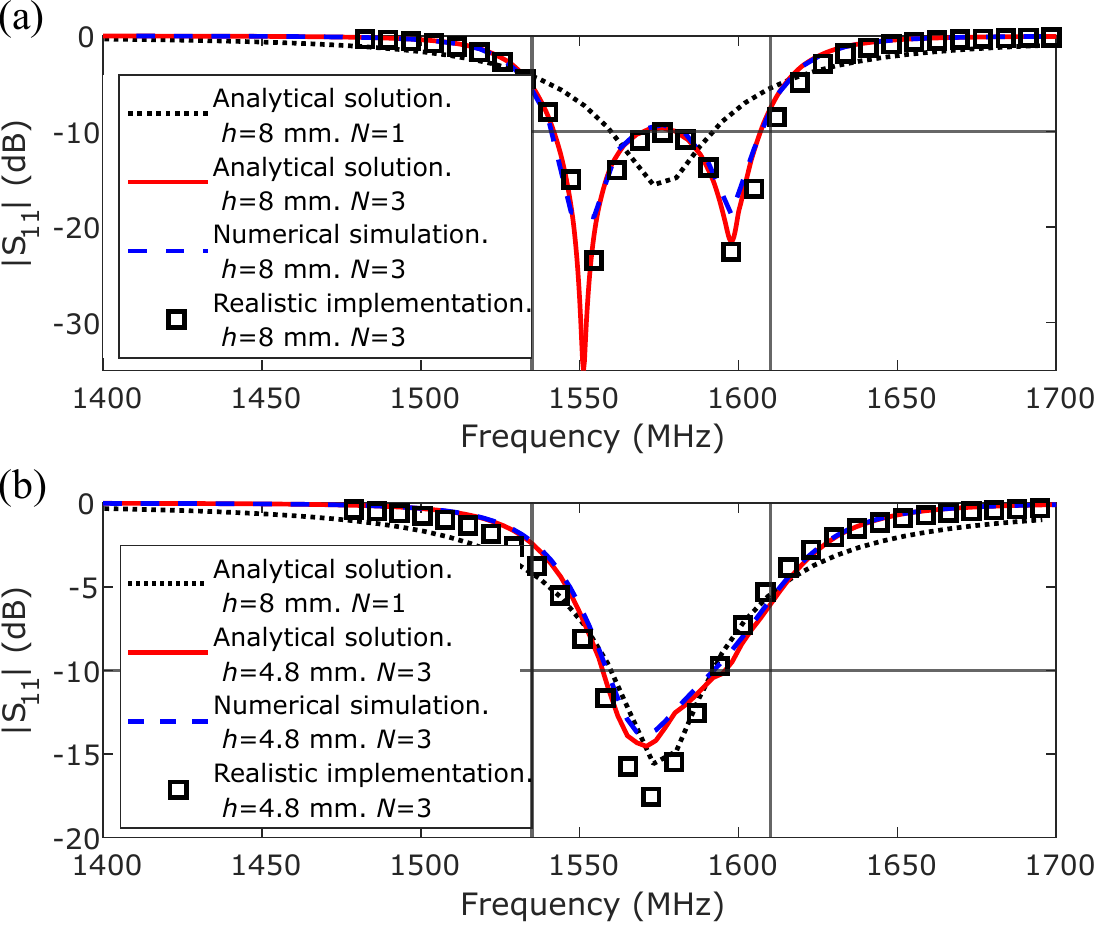}
	\caption{Analytically and numerically calculated frequency dependencies of the reflection coefficient $|S_{11}|$ for two configurations of a synthesized patch antenna with three impedance surfaces ($N=3$), compared to the standard antenna ($N=1$). The markers show numerically calculated curves for the corresponding practical implementations based on periodic loads: (a) demonstration of increasing the frequency bandwidth while maintaining a fixed height $h = 8$ mm; (b) demonstration of reducing the resonator height $h$ from 8 mm to 4.8 mm without changing the matching bandwidth.}
\label{compare_plot} \end{figure}

The analytically predicted frequency dependence of $|S_{11}|$ for the standard antenna and for the antenna configuration with $N=3$ (both having a height of $h=8$~mm) are compared in Fig.~\ref{compare_plot}(a). The goal matching level  $|S_{11}| = -10 ~\text{dB}$ is indicated on the same graph by a horizontal line, while vertical lines indicate the boundaries of the GNSS L1 band. One can observe a bandwidth extension from 32 to 63 MHz. For the antenna with $N=3$, the $|S_{11}|$ curve is compared with full-wave numerical simulation results obtained in CST Microwave Studio software (Finite Element Method implemented in Frequency Domain Solver). In the CST model, impedance surfaces are defined using Ohmic Sheet boundary conditions, while the excitation source is a lumped port at a distance of $r_1$ from the antenna axis. As can be seen from the graph, the resulting numerical curve almost coincides with the analytical one showing the correctness of analytical model. 

The matching bandwidth of patch antennas depends linearly on the resonator height \cite{balanis}. By adding impedance surfaces one can also reduce the height of a circular patch while maintaining a fixed matching band and radial dimensions $a$ and $b$. To demonstrate the possibility of such a miniaturization, we synthesized an antenna with $N=3$ impedance surfaces having the same bandwidth of 32 MHz and the same diameter as for the standard antenna, but with a considerably lower height of 4.8~mm. The corresponding optimized antenna parameters are shown in the last  column of Table~\ref{comparison}. A good coincidence of the analytically calculated $|S_{11}|$ curves for the miniaturized and standard antennas can be seen in Fig.~\ref{compare_plot}(b), which confirms the possibility of reducing the height by more than 60\% without shrinking the  bandwidth. Moreover, the analytical curve for the miniaturized antenna is an a good agreement with a numerical one, as shown in the same graph. 
\section{Practical Implementation of Antennas with Impedance Walls}
To build an antenna based on impedance surfaces, at the second (microscopic) step, it is necessary to associate each internal surface with an equivalent periodic structure of vertical loads \cite{dvtat_gps_satellite_surveying} distributed along the angular coordinate $\varphi$ with a subwavelength step and connecting the patch with the bottom metal plane. Moreover, to ensure the equivalence, the step of loads must be much smaller than the corresponding radius of the cylindrical impedance surface, as well as the distance between adjacent impedance surfaces in the radial direction \cite{sipus}. To implement an impedance surface installed in the radiating slot, it is proposed to place horizontal loads with angular periodicity such that they radially connect the patch to the inner edge of the ground plane in the $z=0$ plane.

The loads of each internal impedance surface should be designed to be compatible with the employed thick air substrate of the antenna, i.e. forming a fence of vertical conductors. The conductors may have different shape depending  on the sign of the imaginary part of $Z_{\text{g},n}$. The 
inductive grid impedance corresponds to a periodic structure of inductive loads. Depending on the value of inductance $L_{n}$, the loads may contain solenoids, meanders, or short-circuited microstrip stubs connected in series to metal pins going vertically from the patch to the bottom metal plane.
For small values of $L_{n}$, straight vertical pins by themselves can provide sufficient inductance. In this case, $L_{n}$ depends on the period and thickness of the pins. A capacitive load must contain a capacitive gap in the path of current flowing vertically from the patch to the bottom plane. Possible implementations may include an open-ended microstrip stub or printed plates connected to a small gap in the vertical pin. Also a capacitive load may be produced by 
just one plate connected to the top end of a vertical pin and located in parallel to the patch plane with a small gap. 
The mentioned solenoids, stubs or plates can be placed on the top of the patch (outside the resonator) to simplify their installation, provided that the corresponding pins go through special holes in the patch. With such an external arrangement, the patch can be realized with the printed-circuit board technology on the bottom side of a thin dielectric substrate. The thickness of the substrate should be small compared to $h$. The vertical pins of the loads pass through via holes and connect to the printed plates (or microstrip stubs) located on the top side of the same substrate. This convenient configuration is used in the further numerical simulations. Note that when placed externally, the loads may cause additional radiation. To ensure that this effect is negligible, a comparison of the radiation pattern with the standard antenna is required.

\begin{figure*}[t]
	\centering  
	\includegraphics[width=1\linewidth]{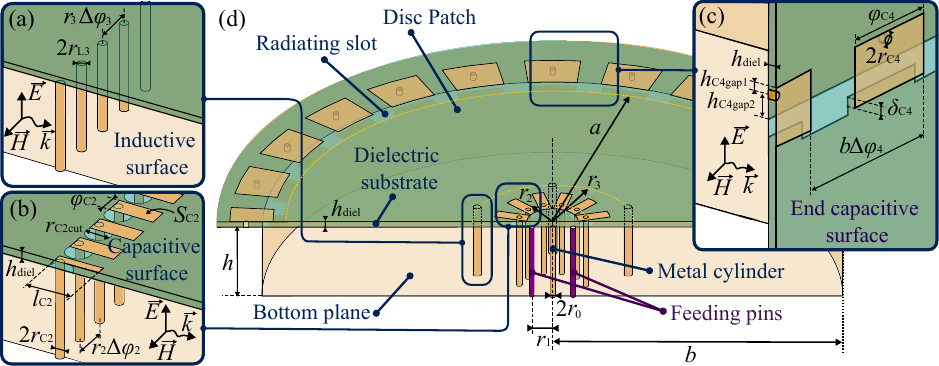}
	\caption{CST models for designing a practical implementation of a circular patch antenna with $N=3$ impedance surfaces: (a) planar periodic structure of the internal inductive impedance surface in a parallel-plane waveguide; (b) planar periodic structure of the internal capacitive impedance surface in a parallel-plane waveguide; (c) planar  periodic structure of the end capacitive impedance surface with a radiating slot on an open flange of a parallel-plane waveguide; (d) the practical implementation of the optimized antenna with $h=8$~mm and an increased bandwidth.}
\label{outer_newver} \end{figure*}

To implement realistic versions of the antennas with $N=3$ (acording to Table~1), inductive loads of the internal walls are implemented as thin metal wires of circular cross-section, while capacitive loads are implemented using pins going through vias connected to capacitive plates on a common dielectric substrate with the patch. The radial capacitive loads periodically distributed along the radiating slot ring are implemented using counter plates, which are continuations of the ground plane and the patch, printed on the same substrate. To determine the specific geometric parameters of the loads for each of the impedance surfaces, CST Microwave Studio software is used as follows. Instead of a periodic structure on a cylinder with radius $r_n$, its flat version is considered (which is acceptable for a small step of loads  $r_n\Delta\varphi_n$ compared to $r_n$). Fig.~\ref{outer_newver}(a,b,c) shows flat versions of periodic structures of inductive internal, capacitive internal, and capacitive radial loads. In the simulation, to select the microstructure parameters of the loads, both an idealized (zero-thickness continuous boundary) and a realistic (discrete) planar versions of the impedance surface are placed in a parallel-plane waveguide. The transmission coefficients for a plane wave incident normal to the surface are numerically calculated. By varying the geometric parameters the same transmission coefficients for the realistic $S_{12}^{\text{real}}$ and idealized $S_{12}^{\text{ideal}}$ surfaces are achieved minimizing $|S_{12}^\text{real} - S_{12}^\text{ideal}|$ in the frequency band.

In order to implement the two examples with $N=3$ analytically synthesized in the previous section, we numerically choose particular structural parameters of the models depicted in Fig.~\ref{outer_newver}(a,b,c). In each case, the
procedure starts with a deliberately small impedance value with its iterative increase via parametric variation. Thus, to increase the grid impedance of the structure shown in Fig.~\ref{outer_newver}(a) assigned to the inductive impedance surface at $r=r_3$, one should reduce radius $r_{\text{L}3}$ of the vertical pins keeping the period $r_3 \Delta\varphi_3$ unchanged. To increase the grid impedance of the structure depicted in Fig.~\ref{outer_newver}(b) assigned to the capacitive impedance surface at $r=r_2$, one should reduce the area $S_{\text{C2}}$ of the printed plates, e.g. by adjusting length $l_{\text{C}2}$.
Finally, to select the proper value of the end capacitive loads (Fig.~\ref{outer_newver}(c)) equivalent to the previously found grid impedance of the continuous impedance surface occupying the area of the radiating slot, one should gradually increase the overlap area of the counter plates. With this aim, parameter $\delta_{\text {C}4}$ could be adjusted. Note that in this case, we consider an open end of a parallel-plane  waveguide with a flat flange rather that a uniform waveguide section. Also, unlike for internal impedance surfaces, the parameters are found by comparing the reflection coefficient of the homogeneous surface ($S_{11}^{\text{ideal}}$) and discrete structure ($S_{11}^{\text{real}}$) at the flange (minimizing the difference $|S_{11}^{\text{real}}-S_{11}^{\text{ideal}}|$ in the frequency band).
Despite the period $r_n \Delta \varphi_n$ of loads could also serve as a variable parameter for approaching the analytically calculated grid impedance $Z_{\text{g},n}$, in the case of a compact antenna with multiple surfaces, one should keep each period as small as possible (at least smaller than the radial distance between the closest adjacent surfaces) to be consistent with the homogenization approximation and avoid excitation of higher-order spatial harmonics within the antenna.

Fig.~\ref{outer_newver}(d) shows a practical implementation of the optimized antenna with $h=8$~mm and $N=3$ (according to Table~\ref{comparison}) analytically predicted to have a bandwidth of 63 MHz. One inductive and two capacitive impedance surfaces are replaced with  periodic structures using the above-described approach.
In the CST model of the antenna the disc patch and capacitive plates of two periodic structures are deposited on two sides of the same dielectric substrate with a thickness of $h_{\text{diel}}=0.508$ mm and a relative dielectric constant of $\varepsilon_{\text{diel}}=3.55$ (corresponding to Rogers RO4003C with a dielectric loss tangent of 0.0027). All conductors of the antenna including the infinite ground plane, bottom plane, patch, central cylinder pins of the inductive impedance surface, and capacitive plates of two capacitive surfaces, are made of copper with a conductivity of $5.96\cdot 10^7$~S/m. The $\text{TM}_{1,1}$ mode is excited by two vertical feeding pins at $r=r_1$ connected to two out-of-phase lumped ports. Practical models for the optimized antenna with $h=4.8$~mm and $N=3$ (see Table~\ref{comparison}), and the standard antenna with $h=8$~mm and $N=1$ are built in a similar way. The numerically determined geometric parameters of all three compared antennas are summarized in Table~\ref{comparison_real}, where $N_\text{C2}$, $N_\text{L3}$ and $N_\text{C4}$ mean the number of loads in the periodic structure of the internal capacitive, internal inductive, as well as capacitive impedance surface in the radiating slot, respectively.

The numerically calculated $|S_{11}|$ frequency curve of the practical antenna with $N=3$ and increased bandwidth is shown by markers in Fig.~\ref{compare_plot}(a). Similar curve for the miniaturized antenna with $N=3$ is shown with markers in Fig.~\ref{compare_plot}(b). It can be seen that for both examples the analytically predicted frequency curves for continuous impedance surfaces are in good agreement with numerically calculated frequency curves for practical periodic structures. Therefore, the dual-step synthesis method, the first step of which is based on the developed analytical model, is verified in two examples. 
\begin{table}
\centering
\caption{Microstructure parameters of the standard antenna ($N=1$), and two optimized antennas with $N=3$}
\begin{tabular}{l c c c}
\hline 
$N$ & 1 & 3 & 3 \\
\hline 
$h$, mm& 8 & 8 & 4.8   \\ \hline
$N_\text{C2}$& - & 12 & 12  \\
$r_\text{C2} $, mm& - & 0.25 & 0.25  \\
$S_\text{C2}$, mm$^2$& - & 12.9 &  21.8    \\
$r_{\text{C}2\text{cut}}$, mm& - & 0.65 &  0.65 \\
$l_\text{C2} $, mm& - & 5.40 & 8.10  \\
$\varphi_{\text{C}2},^\circ$ & - & 26.8 &  26.8 \\
\hline
$N_\text{L3}$& - & 6 & 6   \\
$r_\text{L3}$, mm& - & 0.45 & 0.45 \\
\hline
\multicolumn{4}{c}{$h_{\text{C}4 \text{gap}1}=1.5$ mm, $h_{\text{C}4 \text{gap}2}=1.5$ mm, $r_{\text{C}4}=0.25$ mm} \\
$N_\text{C4}$& 24 & 24 & 24   \\
$\varphi_\text{C4},^\circ$& 10 & 10 & 10 \\
$\delta_\text{C4}$, mm& 0.36 & 0.50 & 1.0 \\

\hline
 Bandwidth, MHz & 32  &  63 & 32   \\
\end{tabular}
\label{comparison_real}
\end{table}
\begin{figure}[t]
	\centering  
	\includegraphics[width=0.95\linewidth]{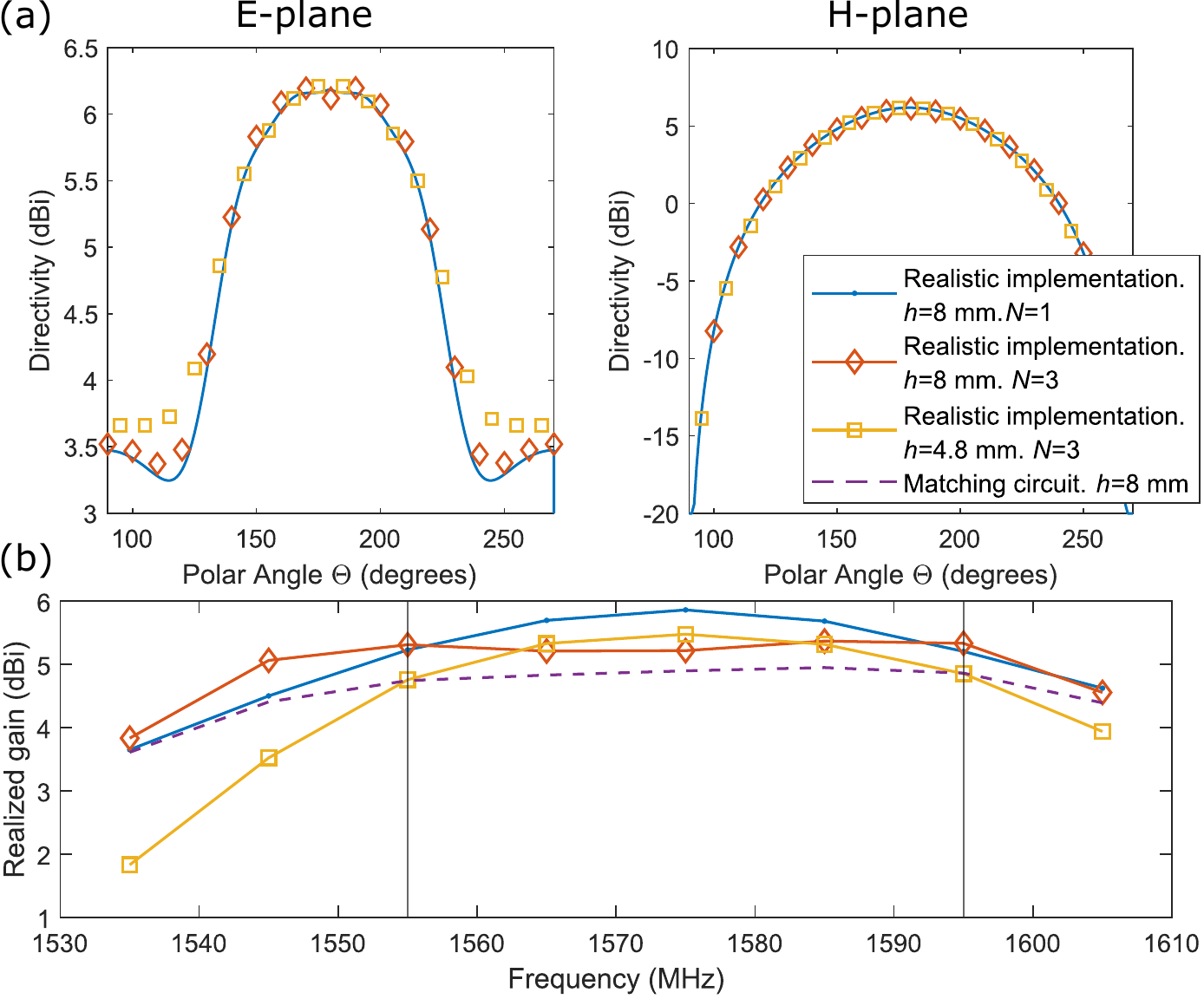}
	\caption{Numerically calculated characteristics of two optimized antennas in their practical implementation: (a) directivity (dBi) in the E- and H-plane at $f\approx f_0$~(1575~MHz); (b) realized gain as a function of frequency.}
\label{DDs_rel_gain} \end{figure}

Fig.~\ref{DDs_rel_gain}(a) shows E-plane and H-plane radiation patterns for three realistic antenna designs: the standard, extended-bandwidth, and miniaturized antennas. It is clearly seen that the shapes of all three radiation patterns coincide well in both planes, which indicates the absence of spurious radiation caused by the presence of realistic capacitive loads in their external  position above the patch. As expected, in all three cases, the far field is formed solely by the radiation from the annular slot. Fig.~\ref{DDs_rel_gain}(b) shows the frequency dependence of the realized gain calculated for three realistic antenna designs. The standard antenna  has the highest realized gain at the central frequency showing, however, pronounced drops by $\approx2.2$~dB at the edges of the band due to impedance mismatch.
A similar behavior is shown by the miniaturized antenna, though its peak realized gain is smaller by 0.4 dB than for the standard antenna. The realized gain of the  extended-bandwidth antenna is lower at the central frequency, but higher near the edges of the band. As one can see, having a comparable averaged realized gain within the band, the latter antenna has a more uniform frequency behavior being more appropriate for transmitting signals without distortion. Reduction of the realized gain for both antennas with $N-3$ is explained  by increased ohmic losses in the conductors of the periodic structures. Summarizing the above results, the proposed impedance surface technique can extend the bandwidth or reduce the height of compact circular patch antennas without deteriorating their radiation patterns at the cost of a reduced radiation efficiency on the order of 0.5 dB.

Since increased dissipation losses is a common drawback of broadband impedance matching techniques, a numerical comparison of the obtained results is performed with the state-of-the-art method of multi-element matching circuits. With this aim, the standard antenna is equipped with a specially synthesised distributed microstrip circuit connected to the inputs of the feeding rods. The elements of the circuit are printed on a separate dielectric substrate situated  below the bottom plane of the resonator. The circuit consists of two resonators connected through a transmission-line section, and a quarter-wave transformer.
Each resonator includes a short-circuited and an open stub both designed in the parallel-type microstrip topology. The matching circuit is designed as follows. First, a second-order Chebyshev filter based on lumped elements is synthesized with the load being the previously calculated input impedance of the standard antenna at the feeding pin. The goal of the filter synthesis is to realize the same bandwidth and the same slope of the frequency response as for the extended-bandwidth antenna with $N=3$.
Second, each of the required lumped elements of the filter is replaced with an equivalent distributed one. Finally, all elements of the filter are implemented as microstrip line sections realized on 0.508-mm-thick Rogers RO4003C substrate and combined with the standard antenna model in CST. The numerically calculated realized gain is shown with a dashed curve in Fig.~\ref{DDs_rel_gain}(b). As can be seen the state-of-the-art methods allows expanding the frequency band, but provides a 0.6 dB lower efficiency and realized gain averaged in the GNSS L1 band than the proposed impedance surface technique. Moreover, the synthesized matching circuit has linear dimensions of 30 $\times $ 21 mm, which is comparable to the dimensions of the patch radiator. Advantageously, the proposed impedance surface technique employs the space inside the antenna consuming no additional space outside of the radiator. In particular, the back side of the bottom plane in our technique is free to accommodate other electronic components of a receiver.

\section{Conclusion}
In this paper, a novel technique for expanding the impedance matching bandwidth and miniaturization of circular patch antennas without radiation pattern distortion has been proposed.  The technique utilizes an optimal set of internal coaxial cylindrical impedance surfaces. An analytical model was developed to calculate the input impedance of a patch antenna with an arbitrary number of impedance surfaces. The model, which was verified via numerical simulations, facilitates solving both analysis and synthesis problems. 

Two examples of improvement in comparison to a standard circular patch antenna were demonstrated, i.e. an antenna with an expanded matching frequency band at the same dimensions, as well as a miniaturized antenna with the same bandwidth. Both antennas were designed by systematically replacing the idealized impedance surfaces with equivalent periodic structures of loads. 

The numerical simulation results showed the possibility to increase the matching bandwidth from 32 MHz to 68 MHz for the same patch diameter of 54 mm and a height of 8 mm in the GNSS L1 band using three impedance surfaces. In comparison to the state-of-the-art method of microstrip matching circuits, the proposed technique does not consume any additional space outside of the radiator, saving space on the printed circuit board for electronic components of a receiver. Moreover, for the same materials used and the same frequency response, the proposed method provides a 0.6-dB-higher radiation efficiency averaged within the band. This advantage is achieved thanks to less confined electric currents flowing in the periodic structures of loads compared to currents in resonant microstrip stubs of broadband matching circuits. The other example demonstrates the possibility of reducing the antenna height from 8 to 4.8 mm while keeping the same frequency response and radiation pattern. As follows from the presented results, the proposed technique can be considered as an effective alternative to conventional microstrip matching circuits. 

Thanks to the developed analytical model, the proposed approach opens new possibilities for precise tailoring the frequency response of compact patch antennas. Apart from increasing the bandwidth and miniaturization, in future work, the developed model combined with nonlinear optimization methods, can be used for prototyping antennas with multiband, higher-order bandpass, notch-filter, and other complex characteristics.

\appendices
\section{}
Let us consider the calculation of $Y_{\text{rad}}$. Due to the equivalence principle, the antenna radiates into the upper half-space like a ring of magnetic current on an infinite metal screen.
Distribution of magnetic current over a gap with a small width compared to the radius ($|b-a| \ll |a+b|/2$) for a mode with azimuthal index $m$ can be taken as $M_{\text{out} \varphi} =  u_0 {{a+b}\over{2 \Delta}} {{e^{-i m \varphi}}\over{r}}$, where $u_0$ is the current amplitude, and the width of the radiating slot is designated as $\Delta = b - a$. The radiation admittance of the given radiator can be represented in the spectral form as the sum of complex admittances corresponding to the excitation of the TE and TM waves in the upper half-space: 
\begin{gather*}
Y_\text{rad}=Y_\text{rad}^{\text{TM}}+ Y_\text{rad}^{\text{TE}}; \\ 
Y_\text{rad}^{\text{TM}} = 2 \pi \int_0^{+\infty}  Y^{\text{TM}} {{a^2}\over{k_r \Delta^2}} (J_{m}(k_r b) - J_{m}(k_r a))^2 dk_r;
\end{gather*}
\begin{gather*}
Y_\text{rad}^{\text{TE}} = 2 \pi \int_0^{+\infty} \frac{m^2 a^2}{\Delta^2} Y^{\text{TE}} I^2(k_r, b, a) \frac{dk_r}{k_r},
\end{gather*}
where $I(k_r, b, a) = \int_{a}^{b} \frac{J_m(k_r r)}{r} d r$ is determined by numerical integration, $k_r$ --- radial component of the wave vector; $J_m$ --- Bessel function of first kind with the order $m$, while $Y^\text{TM} = \frac{\omega 
\varepsilon \varepsilon_0}{\sqrt{k^2-k_r^2}}$ and $Y^\text{TE} = \frac{\sqrt{k^2 - k_r^2}}{\omega \mu \mu_0}$ --- characteristic admittances for the TM and TE waves, respectively.  

Let us consider the admittance matrices $[\text{Y}]_{\text{wg}}$ for a section of a radial waveguide limited by radii $r_{n-1}$ and $r_n$. According to the principle of equivalence, in the single-mode approximation, azimuthal magnetic currents flow uniformly along the $z$ axis at the inner and outer boundaries of the cavity. Using expressions for the fields they create in a radial waveguide, one can obtain \cite{sipus}:

\begin{equation}    
[\text{Y}]_{\text{wg}, n} = \begin{bmatrix}
\frac{\xi(r_{n-1}, r_{n})}{\nu(r_{n-1}, r_n)} & \frac{\xi(r_{n-1}, r_{n-1})}{\nu(r_{n-1}, r_{n})} \\
\frac{\xi(r_{n}, r_{n})}{\nu(r_{n-1}, r_{n})} & \frac{\xi(r_{n}, r_{n-1})}{\nu(r_{n-1}, r_{n})}
\end{bmatrix};
\label{Y_waveguide} 
\end{equation}
\begin{equation*} 
\begin{gathered}
\xi(x, y)  = 
(J_m' (k x)  Y_m(k y) - J_m (k y)   Y_m'(k x)) {{ 2  \pi i x}\over{W_0 h}};  \\
\nu(x, y) = J_m (k x) Y_m (k y) - J_m (k y) Y_m (k x),
\end{gathered} 
\end{equation*}
where $W_0 = 120\pi$ Ohm --- characteristic impedance of free space.

Each parallel-connected load at $r=r_n$, corresponding to an impedance surface with a grid impedance $Z_{\text{g},n}$, in a radial waveguide is associated with a transmission matrix: $[\text{A}]_{\text{g},n} =  \bigl[ \begin{smallmatrix}1 & 0\\ Y_{\text{g},n} (2 \pi r_n)/h  & 1\end{smallmatrix}\bigr]$. If the loads forming the impedance surface contain capacitive loads, then $Z_{\text{g},n}=1/i\omega C_n$, and if they do not (i.e., together they form a grid of parallel thin wires), or --- contain inductive loads, then $Z_{\text{g},n}=i\omega L_n$, where $C_n$ and $ L_n$ --- averaged grid capacitance and inductance of the impedance surface, respectively.

Let us consider the admittance matrix calculation for  the end two-port network $[\text{Y}]_\text{end}$. For the calculation, it is necessary to consider the excitation of a radial waveguide section by a magnetic current $M_{\text{up},\varphi}$ located within the radiating annular slot on the upper metal surface in the plane of the patch. The magnetic current is equal to the current $M_{\text{out},\varphi}$ taken with the opposite sign. Let us find the electric field $\vec{E}_\perp$ transverse to the $z$ axis as a sum of the eigenvector functions $\vec{e}_q$ depending on the geometry of the end cavity, while the transverse magnetic field $\vec {H}_\perp$ --- as a sum of eigenvector functions $\vec{h}_q$.
The mathematical basis for this calculation is presented in \cite{felcen_marcuvitz}. To calculate the spatial distributions of $\vec{e}_q$ and $\vec{h}_q$, we write auxiliary scalar basis functions $\Phi_q$ and $\Psi_q$ for the TM and TE waves, respectively. In contrast to the upper half-space, the magnetic current ring inside the end segment of the radial waveguide creates a field with a discrete spatial spectrum. When expanding into a series, we use the summation index $q$. Scalar basis functions satisfy the two-dimensional Helmholtz equation in coordinates transverse to the $z$ axis $\nabla_\perp^2 \Phi_q + k_r^2 \Phi_q=0$ with boundary conditions on the perfectly conducting walls of the resonator $\Phi_q = 0$ for TM waves and $\frac{\partial}{\partial r} \Psi_q = 0$ for TE waves, where $k_r$ is the radial component of the wave vector.
One can write a scalar basis function and a characteristic equation for $k_{rq}$ --- the radial component of the wave vector for an eigenmode with radial order $q$. For TM waves:
\begin{gather*}
\Phi_q(r) = A^\text{TM} \phi_q(r)  = \\ = A^\text{TM} \left(J_m(k_{rq} r) - \frac{J_m(k_{rq} r_N)}{Y_m(k_{rq} r_N)} Y_m(k_{rq} r)\right) \frac{1}{\sqrt{2 \pi}} e^{-i m \varphi},
\end{gather*}
where $k_{rq}$ satisfies the characteristic equation:
\begin{gather*}
J_m(k_{rq} b) Y_m(k_{rq} r_N) - J_m(k_{rq} r_N) Y_m(k_{rq} b) = 0. 
\end{gather*}
For TE waves:
\begin{gather*}
\Psi_q(r) = A^\text{TE} \psi_q(r) = \\ = A^\text{TE} \left(J_m(k_{rq} r) - \frac{J_m'(k_{rq} r_N)}{Y_m'(k_{rq} r_N)} Y_m(k_{rq} r) \right) \frac{1}{\sqrt{2 \pi}} e^{-i m \varphi}, 
\end{gather*}
where $k_{rq}$ satisfies the characteristic equation:
\begin{gather*}
J_m'(k_{rq} b) Y_m'(k_{rq} r_N) - J_m'(k_{rq} r_N) Y_m'(k_{rq} b) = 0.
\end{gather*}
In the above expressions, due to the orthonormal basis of the eigenmodes:
\begin{gather*}
A^\text{TM} = \sqrt{2/({b^2 (\phi_q'(k_{rq} b) )^2 - r_N^2 (\phi_q'(k_{rq} r_N) )^2})}; \\ 
A^\text{TE} = \sqrt{2/( r_N^2 \psi_q''(k_{rq} r_N) \psi_q(k_r r_N)  - b^2 \psi_q''(k_{rq} b) \psi_q(k_{rq} b)}).
\end{gather*} 
The basis vector functions then can be written as:
\begin{gather*}
\vec{e}_q^\text{TM} = -\frac{\nabla_\perp \Phi_q}{k_{rq}}; ~~~~~~~~
\vec{h}_q^\text{TM} = -\vec{z}_0 \times \frac{\nabla_\perp \Phi_q}{k_{rq}}; \\
\vec{e}_q^\text{TE} =  - \frac{\nabla_\perp \Psi_q}{k_{rq}} \times \vec{z}_0; ~~~~~~~~
\vec{h}_q^\text{TE} = -\frac{\nabla_\perp \Psi_q}{k_{rq}}.
\end{gather*}
Considering the boundary conditions at the perfectly conducting boundary of the end block in the $z=h$ plane, the total transverse components of the electric and magnetic field inside the end cavity can be written as a superposition of TM and TE waves:
\begin{equation}
\begin{gathered}
\vec{E}_\perp = \sum_{q=1}^{+\infty} V_q^\text{TM} \vec{e}_q^\text{TM} ( e^{-i k_{zq} z}  - e^{i k_{zq} (z-2h)} ) + \\ + \sum_{q=1}^{+\infty} V_q^\text{TE} \vec{e}_q^\text{TE} ( e^{-i k_{zq} z}  - e^{i k_{zq} (z-2h)} ); \\
\vec{H}_\perp = \sum_{q=1}^{+\infty} Y^\text{TM} V_q^\text{TM} \vec{h}_{q}^\text{TM}  (e^{-i k_{zq} z}  + e^{i k_{zq} (z-2h)})  + \\
+ \sum_{q=1}^{+\infty} Y^\text{TE} V_q^\text{TE} \vec{h}_{q}^\text{TE} ( e^{-i k_{zq} z}  + e^{i k_{zq} (z-2h)} ).
\label{sum_field}
\end{gathered}
\end{equation}
The amplitudes $V_q^\text{TM/TE}$ for each of the harmonics of a discrete spatial spectrum can be calculated from the boundary conditions on the antenna patch containing the radiating slot, using the property of orthogonality and normalization of eigenfunctions. 
\begin{equation*}
\begin{gathered}
V_{q}^\text{TM/TE}  = - \frac{\vec{z}_0 \int_a^b \int_0^{2\pi} [\vec{e}_{q}^{*\text{TM/TE}} \times \vec{M}_\text{S}] d\varphi dr}{( 1  - e^{-2i k_{zq} h} )}. 
\label{harm_amp}
\end{gathered}
\end{equation*}
After calculating the tangential component of the electric field (and, accordingly, the magnetic current density) on the surface $r=r_{N}$ and on the radiating slot ($z=0$, $a<r<b$), it is possible to calculate all components of the admittance matrix for the patch antenna end block $[\text{Y}]_\text{end}$.
\begin{equation}
\begin{gathered}
Y_{\text{end}, 11} = \xi(r_N, b) / \nu(r_N, b); \\
Y_{\text{end}, 21}^* = Y_{\text{end}, 12}^* = -\frac{ 2\pi}{u_1 u_0^*} \int_0^h (\vec{\varphi}_0 \cdot \vec{H}_\perp^{*}) M_{\text{in} \varphi}  dz;  \\
Y_{\text{end}, 22}^* = -\frac{2 \pi}{u_0 u_0^*} \int_a^b (\vec{\varphi}_0 \cdot \vec{H}_\perp^{*}) M_{\text{up} \varphi} r dr, 
\label{Y_end_matrix}
\end{gathered}
\end{equation}
where $u_1 = M_{\text{in} \phi} \cdot h $ is the amplitude of the magnetic current located at the boundary of the end block with radius $r_N$. It should be noted that to calculate the elements of the admittance matrix using the formulas (\ref{Y_end_matrix}), for certain $r_N$ and $b$ it is necessary to determine the number of modes $q_{\text{max}}$, which is sufficient to take into account to achieve convergence in the expressions (\ref{sum_field}). This operation is performed numerically, as well as for the calculation of integrals in expressions (\ref{Y_end_matrix}).

\nocite{*}

\bibliographystyle{IEEEtran}
\bibliography{aipsamp}

\end{document}